\documentclass[manuscript]{aastex62}
\usepackage{amsmath}

\usepackage{fourier} 
\usepackage{makecell}

\hypersetup{linkcolor=red,citecolor=blue,filecolor=blue,urlcolor=blue}

\newcommand{\vhz}{$\mbox{V}^2\;\mbox{Hz}^{-1}~$}
\newcommand{\wmhz}{$\mbox{W}\;\mbox{m}^{-2}\;\mbox{Hz}^{-1}~$}
\received{}
\revised{}
\accepted{}

\shorttitle{Spectral Analysis of Solar Radio Type III Bursts}
\shortauthors{Sasikumar Raja et al.}

\begin{document}

\title{Spectral Analysis of Solar Radio Type III Bursts from 20 kHz to 410 MHz}

\correspondingauthor{K. Sasikumar Raja}
\email{sasikumar.raja@iiap.res.in; sasikumarraja@gmail.com}

\author[0000-0002-1192-1804]{K. Sasikumar Raja}
\affil{LESIA, Observatoire de Paris, Universit\'e PSL, CNRS, Sorbonne
Universit\'e, Universit\'e de Paris, 5 place Jules Janssen, 92195 Meudon,
France.}
\affil{Indian Institute of Astrophysics, II Block, Koramangala, Bangalore-560 034, India}

\author[0000-0001-6172-5062]{Milan Maksimovic}
\affiliation{LESIA, Observatoire de Paris, Universit\'e PSL, CNRS, Sorbonne
Universit\'e, Universit\'e de Paris, 5 place Jules Janssen, 92195 Meudon,
France.}

\author[0000-0002-8078-0902]{Eduard P. Kontar}
\affiliation{School of Physics and Astronomy, University of Glasgow, G12 8QQ Glasgow, UK.} 

\author{Xavier Bonnin}

\affiliation{LESIA, Observatoire de Paris, Universit\'e PSL, CNRS, Sorbonne
Universit\'e, Universit\'e de Paris, 5 place Jules Janssen, 92195 Meudon,
France.}

\author{Philippe Zarka}

\affiliation{LESIA, Observatoire de Paris, Universit\'e PSL, CNRS, Sorbonne
Universit\'e, Universit\'e de Paris, 5 place Jules Janssen, 92195 Meudon,
France.}
\affiliation{Unit\'e scientifique de Nan\c{c}ay, Observatoire de Paris, CNRS, PSL, Universit\'e d'Orl\'eans/OSUC, Nan\c{c}ay, France.}
\author{Laurent Lamy}
\affiliation{LESIA, Observatoire de Paris, Universit\'e PSL, CNRS, Sorbonne
Universit\'e, Universit\'e de Paris, 5 place Jules Janssen, 92195 Meudon,
France.}
\affiliation{Unit\'e scientifique de Nan\c{c}ay, Observatoire de Paris, CNRS, PSL, Universit\'e d'Orl\'eans/OSUC, Nan\c{c}ay, France.}
\affiliation{Aix Marseille Univ, CNRS, CNES, LAM, Marseille, France.}
\author[0000-0002-6287-3494]{Hamish Reid}
\affiliation{Department of Space \& Climate Physics, University College London, UK.}

\author[0000-0002-6872-3630]{Nicole Vilmer}
\affiliation{LESIA, Observatoire de Paris, Universit\'e PSL, CNRS, Sorbonne
Universit\'e, Universit\'e de Paris, 5 place Jules Janssen, 92195 Meudon,
France.}
\affiliation{Unit\'e scientifique de Nan\c{c}ay, Observatoire de Paris, CNRS, PSL, Universit\'e d'Orl\'eans/OSUC, Nan\c{c}ay, France.}

\author{Alain Lecacheux}
\affiliation{LESIA, Observatoire de Paris, Universit\'e PSL, CNRS, Sorbonne
Universit\'e, Universit\'e de Paris, 5 place Jules Janssen, 92195 Meudon,
France.}

\author[0000-0001-6185-3945]{Vratislav Krupar}
\affiliation{Goddard Planetary Heliophysics Institute, University of Maryland, Baltimore County, Baltimore, MD 21250, USA.}
\affiliation{Heliophysics Science Division, NASA Goddard Space Flight Center, Greenbelt, MD 20771, USA.}
\author{Baptiste Cecconi}
\affiliation{LESIA, Observatoire de Paris, Universit\'e PSL, CNRS, Sorbonne
Universit\'e, Universit\'e de Paris, 5 place Jules Janssen, 92195 Meudon,
France.}
\affiliation{Unit\'e scientifique de Nan\c{c}ay, Observatoire de Paris, CNRS, PSL, Universit\'e d'Orl\'eans/OSUC, Nan\c{c}ay, France.}
\author{Lahmiti Nora}
\affiliation{LESIA, Observatoire de Paris, Universit\'e PSL, CNRS, Sorbonne
Universit\'e, Universit\'e de Paris, 5 place Jules Janssen, 92195 Meudon,
France.}

\author{Laurent Denis}
\affiliation{Unit\'e scientifique de Nan\c{c}ay, Observatoire de Paris, CNRS, PSL, Universit\'e d'Orl\'eans/OSUC, Nan\c{c}ay, France.}




\begin{abstract}
We present the statistical analysis of the spectral response of solar radio type III bursts over the wide frequency range between 20 kHz and 410 MHz. For this purpose, we have used observations that were carried out using both spaced-based (Wind/Waves) and ground-based (Nan\c{c}ay Decameter Array and Nan\c{c}ay Radioheliograph) facilities. In order to compare the flux densities observed by the different instruments, we have carefully calibrated the data and displayed them in Solar Flux Units.  
The main result of our study is that type III bursts, in the metric to hectometric wavelength range, statistically exhibit a clear maximum of their median radio flux density around 2 MHz. Although this result was already reported by inspecting the spectral profiles of type III bursts in the frequency range 20 kHz - 20 MHz, our study extends such analysis for the first time to metric radio frequencies (i.e., from 20 kHz to 410 MHz) and confirms the maximum spectral response around 2 MHz. In addition, using a simple empirical model we show that the median radio flux $S$ of the studied dataset obeys the polynomial form $Y = 0.04 X^3 - 1.63 X^2 + 16.30 X -41.24$, with $X=\ln{(F_\text{MHz})}$ and with $Y=\ln{(S_\text{SFU})}$. Using the Sittler and Guhathakurtha model for coronal streamers \citep{Sit1999}, we have found that maximum of radio power falls therefore in the range 4 to 10 $R_{\odot}$, depending on whether the type III emissions are assumed to be at the fundamental or the harmonic.

\end{abstract}



\section{Introduction} \label{sec:intro}

Solar type III radio bursts are produced by electron beams that are propagating along open magnetic field lines in the corona and interplanetary medium (IPM). Among the other radio bursts classified by \citet{Wild1950, Wild1967}, type IIIs are the most intense, fast drifting (0.1 - 0.5 $c$, where c is the speed of light) and frequently observed bursts. It is widely accepted that type III bursts originate via a plasma emission mechanism \citep{Ginzburg1958} in which the fast electrons form an unstable velocity distribution leading to a bump-on-tail instability \citep{1949PhRv...75.1864B} which excites Langmuir waves at the local plasma frequency ($f_p=8.98 \times \sqrt{N_e}$, where $f_p$ is the plasma frequency in kHz and $N_e$ is the electron density in $\text{cm}^{-3}$). These Langmuir waves are then transformed into electromagnetic radiations by plasma emission mechanisms \citep{Ginzburg1958}. For instance, the coalescence of Langmuir waves and low-frequency ion-sound waves can produce radiation at the electron plasma frequency ($f_p$) called Fundamental or F-emission, whereas the coalescence of two Langmuir waves can lead to radiation at the frequency $\approx 2f_p$ called Harmonic or H-emission \citep{Ste1974,Mel1987,Rob1994,The2018}. 
As the electron density and therefore the plasma frequency decreases radially outwards, type III bursts drift from high frequencies ($f_p \approx 1$ GHz) in the low corona down to the local plasma frequency of the observer, $f_p \approx 20$ kHz at 1 Astronomical Unit (AU). 
It has been known for a long that Type IIIs are extremely variable, 
both in radio flux density and in observing frequency range \citep{Web1978,Dulk1984}. A statistical study of type III bursts \citep{Bon2008b} observed by the Wind/Waves instrument \citep{Bou1995} 
have also found a maximum response of type III bursts at around 1 MHz previously reported by \cite{Web1978}.
More recently, using the S/Waves instruments onboard STEREO twin spacecraft \citep{Bou2008, Kai2008}, \citet{Krupar2014} have confirmed the statistical maximum of flux density around $1$ MHz. In this article, we have studied the spectral response of the type III bursts over a much wider bandwidth (i.e., 20 kHz - 410 MHz) and found that the maximum spectral response lies between 1 - 2 MHz. Furthermore this study confirms that there are no other local maximum of the radio flux in the analyzed frequency bandwidth.

Several physical ingredients (e.g., electron density, electron beam speed, and beam density) and mechanisms determine the radio flux density of a Type III burst and its variation with frequency. Variation of the flux density can be related to the total energy and beam density of the interplanetary energetic flare electrons \citep{Dulk1984}.  In addition, 
the local density fluctuations and/or turbulence can affect the energy of the beam 
and Langmuir waves \citep{1976JPSJ...41.1757N,1985SoPh...96..181M}.

In their study \citet{Krupar2014} have proposed a simple model in which the radio emission is at harmonic and saturated with a radio flux density proportional to the energy of Langmuir waves \citep{Mel1980}. If the latter is a simple function of the electron density and energy distribution, then the notable change of density gradient of the corona/solar wind in the 2 to 5 $R_{\odot}$ radial range \citep{Sit1999} creates a maximum of the Type III radio flux densities at around 1 MHz. However, the latter model does not fit perfectly the median flux density radial profile from \citet{Krupar2014}. 

In the present paper, we have extended for the first time the studies by \citet{Bon2008b, Lecacheux2000} and \citet{Krupar2014} to metric-kilometric radio frequencies by combining the radio observations from space (Wind/Waves) with those from the ground-based Nan\c{c}ay Decameter Array \citep[NDA;][]{Boischot1980,Lecacheux2000} and Nan\c{c}ay Radioheliograph \citep[NRH;][]{Ker1997, Mer2006} that are located at the Station de Radioastronomie de Nan\c{c}ay, France (latitude $47^{\circ} 23\arcmin$ N; longitude $2^{\circ} 12\arcmin$ E; altitude 150 m). By doing so, we extend the study of variation of the flux density with the frequency of type III bursts up to 410 MHz. Consequently, we provide new constraints on the physical mechanism which defines the radio flux density of Type III bursts and its variation with frequency.

In Section \ref{sec:observations} of this article, we describe the various radio instruments that are used for this study. Also, we focus on the techniques that are used for calibrating the observed voltage power spectral densities into radio density flux in physical units. In Section \ref{sec:results} the obtained results are presented along with the interpretations and discussions. Finally in Section \ref{sec:summary}, we provide a summary and conclusions.

\section{Instruments, Observations and data analysis}\label{sec:observations}

In this article, we have studied the spectral profiles of isolated type III bursts observed using different instruments briefly presented here. Based on the observing periods, locations and instruments, we have categorized them into three cases. The first one, henceforth dataset-I, contains type III bursts which are observed using the WAVES instrument \citep{Bou1995}, onboard the Wind spacecraft located at L1, irrespective of their longitudinal source location in the corona. This can even include type IIIs generated by electron beams originating on the opposite side of the Sun, as seen from L1. The second dataset, henceforth dataset-II, contains type III bursts which are simultaneously observed by Wind/WAVES at L1 and by the ground-based Nan\c{c}ay Decameter Array \citep{Boischot1980}. Finally, the third dataset,  henceforth dataset-III, contains imaging observations of type III bursts that are carried out using the NRH \citep{Reid2017}. The details of the three datasets, including the frequency ranges of observations, are summarized in Table \ref{tab:datasets}. The dataset-I consists of 1434 type III bursts that were observed during 1995-2009. The dataset-II consists of 115 type III bursts that were observed during 2013-2014. The dataset-III consists of 218 type III bursts observed during 2002 - 2011. In this case, Wind/WAVES, NDA, and NRH are observing type III bursts that are generated on the visible side of the Sun with respect to the line-of-sight (LOS) direction.

 \begin{table}[!ht]
 \centering
    \begin{tabular}{|c|c|c|c|c|}
      \hline \hline
      Dataset & Instrument & \makecell{Observed \\ frequency} & \makecell{Observed \\ period} & \makecell{No. of \\ type III bursts \\ observed} \\
      \hline \hline
    Dataset-I & \makecell{Wind/Waves \\ (RAD1 and RAD2)} & 20 kHz - 13.825 MHz & 1995 - 2009 & 1434  \\
    \hline
    Dataset-II & \makecell{Wind/Waves \\ (RAD1 and RAD2) \\ and  Nan\c{c}ay Decameter\\ Array}  & \makecell{20 kHz - 13.825 MHz\\ \& \\10 - 80 MHz } & 2013 - 2014 & 115 \\
    \hline
    Dataset-III & Nan\c{c}ay Radioheliograph & 150 - 450 MHz & 2002 - 2011  & 218 \\ 
\hline \hline
    \end{tabular}
\label{tab:datasets}
    \caption{Datasets from Wind/Waves Nancay Decameter Array Nancay Radioheliograph used in the study.}
 \end{table}


In the following sub-section, we briefly present the radio instruments used in this study, together with the irrespective calibration techniques.

\subsection{Wind/Waves}

The Wind/Waves experiment \citep{Bou1995} has three electric dipole antennas; two of them are co-planar and orthogonal wire dipole antennas in the spin-plane, whereas the other is a rigid spin-axis dipole. The longer dipole of Waves experiment ($E_x$) is of $2 \times 50$ m tip-to-tip and the other spin-plane dipole ($E_y$) length is $2 \times 7.5$ m tip-to-tip. The length of spin axis dipole ($E_z$) is $2 \times 6$ m tip-to-tip.
The Wind spacecraft spins at 20 turns per minute (\emph{i.e.} 1 rotation in 3s). Among others, Waves experiment has two radio receiver bands, namely radio receiver band - 1 (RAD1) and radio receiver band - 2 (RAD2). 

RAD1 is designed to operate from 20 kHz to 1040 kHz (sweeping a band-pass filter 3 kHz). RAD2 operates from 1.075 MHz to 13.825 MHz (sweeping a band-pass filter of 20 kHz). Both RAD1 and Rad2 have 256 channels with a sensitivity of $7~\rm nV~Hz^{-{1/2}}$. In this article, we use the observations carried out in S mode, i.e., one receiver is connected to $E_z$ and the other to the sum of $E_z$ and the shorter antenna ($E_y$) signals. The calibration technique we have used is described as follows. 

\begin{figure}[!ht]
\centering
\includegraphics[width=0.9\textwidth]{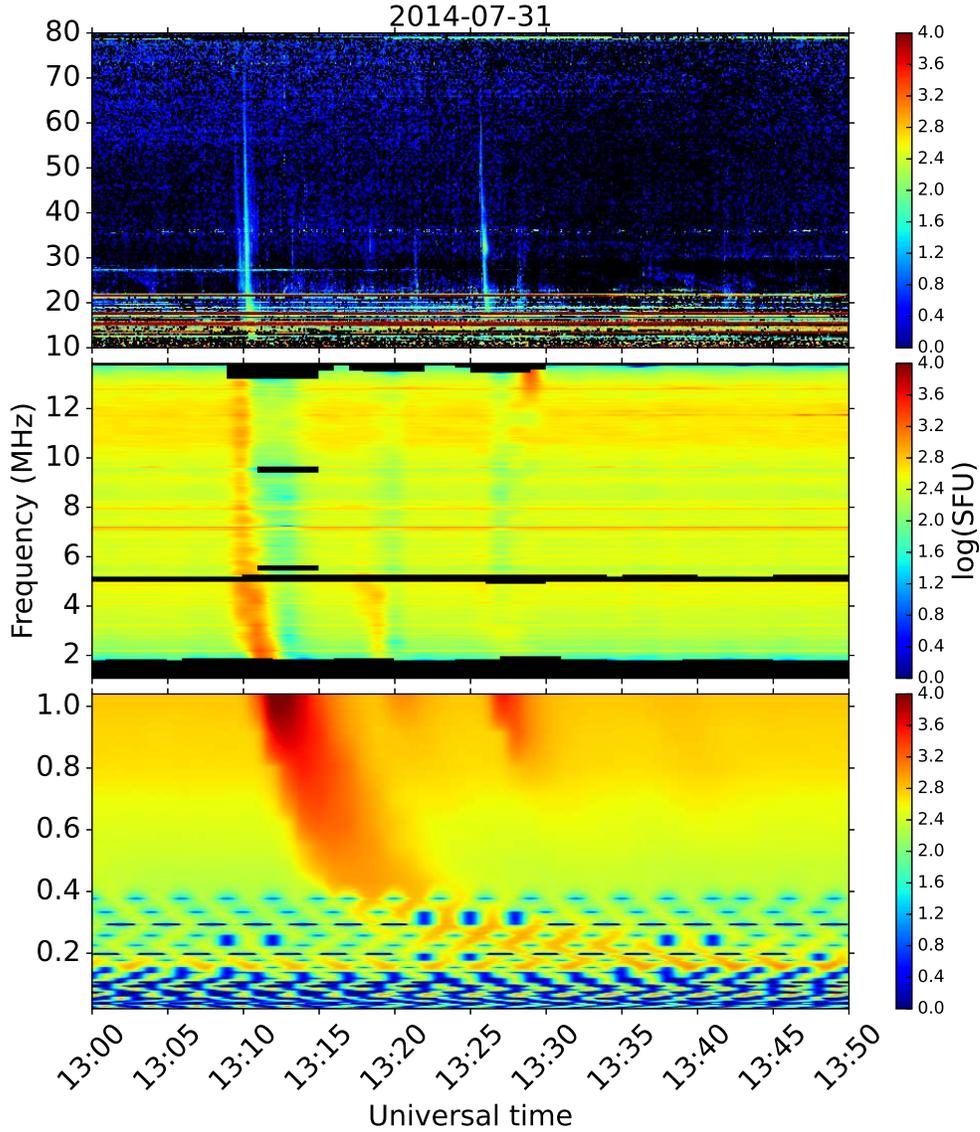}
\caption{Calibrated dynamic spectrogram observed on 31 July 2014. The top panel shows the observation of type III bursts carried out using the Nan\c{c}ay Decameter Array in the frequency range 10-80 MHz. The middle and lower panels show observations carried out using the Wind/Waves instrument from space. The frequency range 20 kHz - 1040 kHz (i.e., lower panel) is observed using RAD1 receiver, and the frequency range 1.075 MHz - 13.825 MHz (i.e., middle panel) is carried out using RAD2 receiver. We note that a type III burst at 13:30 UT is not seen in RAD2 because of missing information. 
}
\label{fig:ds}
\end{figure}

The observed voltage power spectral density $P_{obs}$ (in units of $\text{V}^2\;\text{Hz}^{-1}$) by the RAD1 and RAD2 receivers can be represented as

\begin{equation}
P_{obs}=P_{burst}+P_{rec}+P_{gal},
\end{equation}
where $P_{burst}$ is the voltage power spectral density that corresponds to the radio emission from the Sun which we are interested in, $P_{rec}$ is the receiver noise, and $P_{gal}$ is the voltage power spectral density that corresponds to the galactic background. In order to measure $P_{burst}$, we have to subtract both $P_{rec}$ and $P_{gal}$ from $P_{obs}$. 
We know that the measurements of $P_{rec}$ that was carried out from the ground were contaminated by electromagnetic interference in the laboratory. Therefore, before deployment of the antennas (\emph{i.e.}, from space after the launch), the receiver noise was measured and is subtracted from $P_{obs}$ accordingly.

In order to have the true flux density of the galactic background ($S_{gal}$ in \wmhz), we used the galactic background model of \citet{Nov1978},
\begin{equation}
S_{gal} = S_{g0} \times f^{-0.76}_\text{MHz}e^{-\tau} \times \Omega,
\end{equation}

where $S_{g0} = 1.38 \times 10^{-19}~\mbox{Wm}^{-2}\mbox{Hz}^{-1}\mbox{sr}^{-1}$, $f_\text{MHz}$ is the observed 
frequency in MHz, $\Omega = 8 \pi/3$ sr is the radio beam of the dipole in the short wavelength approximation. \\

We then calibrated the observed flux in \vhz to \wmhz  using the approach by \cite{Zarka2004} and \cite{Zas2011} and using the formula.

\begin{equation}
S_{wg} = (P_{obs}-P_{rec}) \times \frac{S_{gal}}{P^{obs}_{gal}},
\end{equation}

where $S_{wg}$ is the flux density in \wmhz with contribution from the galactic background included and 
$P^{obs}_{gal}$ is the observationally derived galactic background by inspecting large data set. After the calibration, we measured the flux density of the radio burst $S_{burst}$ by subtracting  $S_{gal}$:

\begin{equation}
S_{burst} = S_{wg} - S_{gal}
\end{equation}

\subsection{Nan\c{c}ay Decameter Array}

The Nan\c{c}ay Decameter Array (NDA) observes quasi-daily intense radio bursts from the solar corona over 10-80 MHz \citep{Boischot1980, Lecacheux2000, Lamy2017, Lamy2021}. It is a phased array consisting of 144 helical antennas (with a height of 9 m and diameter of 5 m) divided into two sub-arrays which are sensitive to left-handed (LCP) and right-handed circular polarization (RCP).
Each sub-array (with 72 antennas) is further divided into nine blocks with eight antennas in each group.
The so-called Routine swept-frequency back-end receiver used in this study has been continuously in operation since 1990. It records a full spectrum successively from each polarized sub-array every 0.5s, with 400 frequency channels (when covering 10-80 MHz, this yields a 175 kHz spectral step). It records alternative polarizations from one spectrum to the next. 
This study used Routine raw data expressed in dB of input voltage power spectral density in \vhz, sampled over 8 bits, and calibrated using the pipeline described below. NDA measurements include hourly calibration sequences, during which each block is unplugged from its 8-antenna set and connected to a noise diode of known flux instead. Such calibration sequences can be inverted to calibrate the raw data into absolute flux densities \citep{De1979}. Note that this method does not include the response of the antenna themselves but provides a fair first-order calibration \citep[see][]{Lecacheux2000}.

Each noise diode delivers $41.2 \pm 0.3~dB_{ENR}$, which, summed over the nine blocks of each sub-array, yields a total flux of $50.75 \pm 0.9~ dB_{ENR}$ (L. Denis, personal communication). A calibration sequence decomposes into $4 \times 10$s long intervals, during which the diode signal is sampled with a 0, 10, 20, and 30 dB attenuation. These calibration measurements can thus be inverted to derive the equivalent antenna temperature following the equation

\begin{equation}
T_A \approx T_o \times 10^{dB_{ENR}/10} \; \mbox{K},
\end{equation}

where $T_o$ is the ambient temperature ($\approx 290~ K$). The antenna temperature can then be converted into flux density using

\begin{equation}
S = \frac{2 k_B (T_A-T_{sys})}{A_e}\;\;~\mbox{W}\;\mbox{m}^{-2}\; \mbox{Hz}^{-1}, 
\end{equation}

where, $k_B$ is the Boltzmann constant 
($\simeq 1.38 \times 10^{-23}~ \mbox{JK}^{-1}$), 
$A_e$ is the total effective area modeled by the NDA team, 
and $T_{sys}$ is the system temperature which we estimated for this study to be $\approx 17400 \pm 4000$ K. This calibration pipeline was applied separately for LCP and RCP data. 
The sum of LCP and RCP calibrated flux densities finally provides the total flux density (Stokes parameter I).

\subsection{Nan\c{c}ay Radioheliograph}

The NRH \citep{Ker1997, Mer2006} is a two-dimensional imaging instrument that is capable of observing up to 10 frequencies in the range 150 - 450 MHz \footnote{\url{http://bass2000.obspm.fr/soft\_guide.php}}. The NRH is a T-shaped interferometer with 19 antennas in the east-west direction spread over 3200m and 24 antennas in the north-south direction over 2440m. After the year 2003, the number of antennas was increased to 48. In this work, we used the observations that are carried out at frequencies 164, 236, 327, and 410~MHz. At these frequencies, the flux density of Type III bursts is calculated using routines that generates cleaned images of the burst. The flux density is measured over a square window of size $440 \times 440$ arcsec around the location of maximum radio brightness and over an integration time of 10s. It is worth mentioning that the radio galaxy Cygnus A (an intense radio source with known brightness) is used to calibrate the images. As Cygnus A is an extended source, an analytical model of the source is determined. Then the complex visibility function is derived with an accuracy of a few percent up to the highest spatial frequency of both arrays ($\approx 200~\text{rad}^{-1}$). The instrumental phase and gain fluctuations are about $5^\circ$ and $5\%$, respectively. 

\section{Results and Discussions}\label{sec:results}

\subsection{Peak spectral response of type III bursts at 1.5 to 2 MHz}

\begin{figure}[!ht]
\centerline{\includegraphics[width=0.95\textwidth]{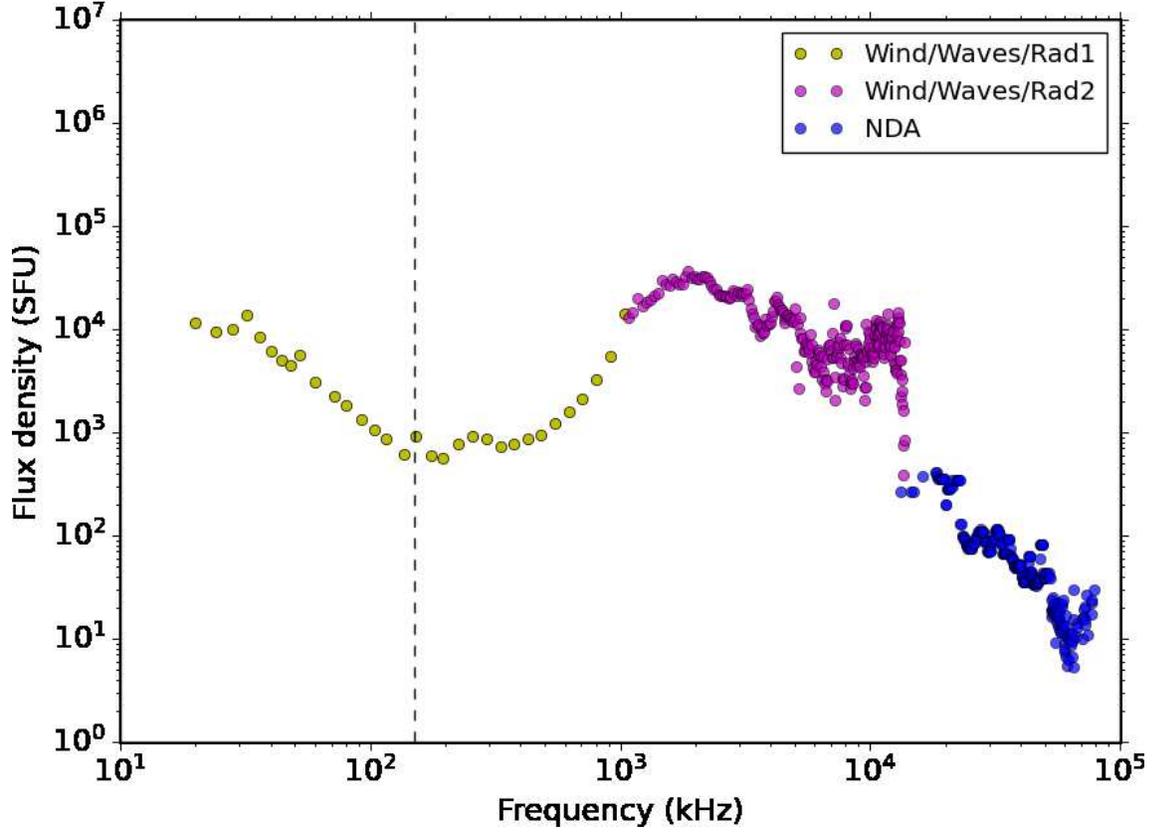}}
\caption{The maximum flux densities of type III burst in different frequency channels that were observed simultaneously using Wind/Waves and NDA are shown for a typical event from dataset-II. The yellow, magenta and blue markers indicate the measurements derived using NDA, Wind/Waves/RAD1, and Wind/Waves/RAD2 instruments. The dotted vertical line delimitates the range below $\approx 150$ kHz where the signal is due to the plasma quasi-thermal noise detected by the antenna.}
\label{fig:peak}
\end{figure}

After calibrating all the events in the three datasets I, II, and III have been inspected to measure the maximum flux densities of the type III bursts for each observing frequency for each event. For instance, Figure~\ref{fig:ds} shows a typical calibrated dynamic spectrogram of Type III burst observed simultaneously by Wind/Waves (20 kHz - 13.825 MHz) and NDA (10 - 80 MHz) on 2014 July 31.
For this event, which belongs to dataset-II, we have determined the peak flux density of the type III burst for every single frequency channel that was recorded during the time interval 13:10 - 13:25 UT. Figure~\ref{fig:peak} shows the distribution of these maxima as a function of the observing frequency. The dotted vertical line indicates the frequency below which the signal corresponds to the electron quasi-thermal noise (QTN) detected by the antenna \citep{Nicole1989}. Considering frequencies above this dotted line, a global maximum flux density at $\approx 2$ MHz can be seen.

\begin{figure}[!ht]
\centerline{\includegraphics[width=0.95\textwidth]{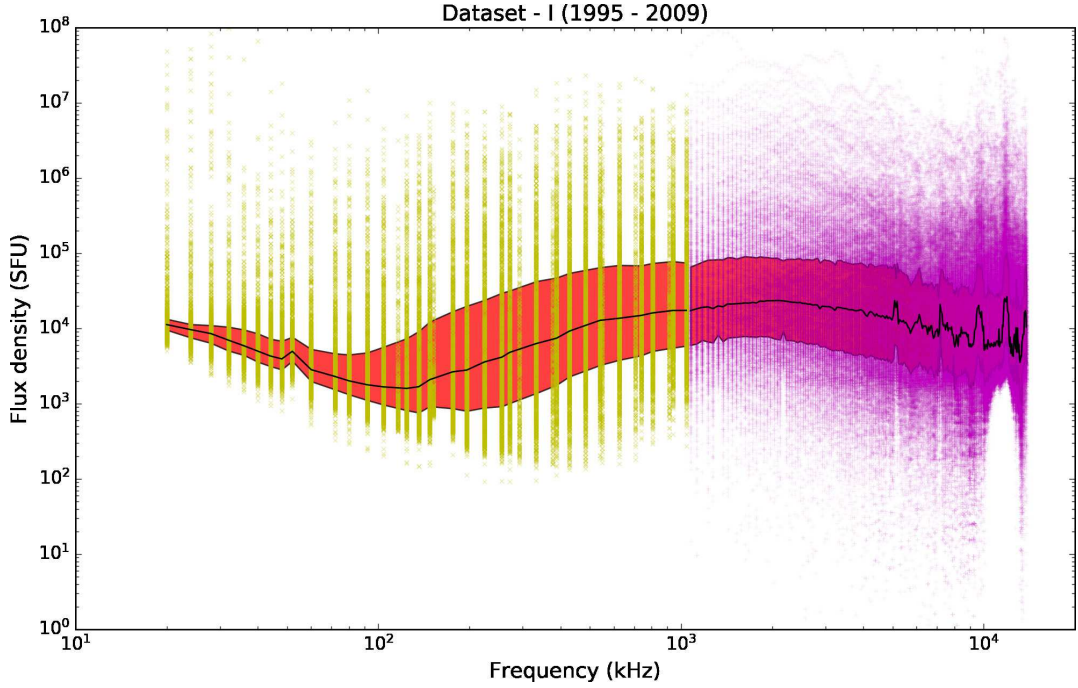}}
\caption{Maximum flux density of type III bursts that were observed during 1995-2009 is shown. The yellow `$\times$' and magenta `$+$' markers indicate the observations of RAD1 and RAD2 receivers, respectively. The black solid curve represents the median (\emph{i.e.}, 50th percentile) of the peak flux densities of the data shown in every frequency channel. Red colored filled region shows the first and third quartiles of the data shown. Note that the galactic contribution is subtracted as described in Section \ref{sec:observations}.}
\label{fig:wind2000}
\end{figure}

Similarly, we have repeated this method and plotted the spectral response of the 1434 type III bursts of dataset-I in Figure~\ref{fig:wind2000}. The yellow `$\times$' and magenta `$+$' markers indicate the measurements derived using RAD1 and RAD2 receivers, respectively. The solid black curve represents, for each observing frequency, the median of the data points for all the events. The filled red color region goes from first to third quartiles. As one can see, we confirm and refine the result already reported by \citet{Bon2008b} and \citet{Krupar2014}. When considering a large enough number of events, the typical type III radio emission exhibits a maximum at 1.5 to 2 MHz. In the studies by \citet{Bon2008b} and \citet{Krupar2014} the maximum is close to 1 MHz, while in the present study, it is closer to 2 MHz. Furthermore, the flux at the spectral peak in Figure~\ref{fig:wind2000}, which is $\approx 2 \times 10^4$ SFU, is almost the same as the maximum flux of $\approx 3 \times 10^4$ reported by  \citet{Krupar2014} on their Figure 10. 

\begin{figure}[!ht]
\centerline{\includegraphics[width=0.95\textwidth]{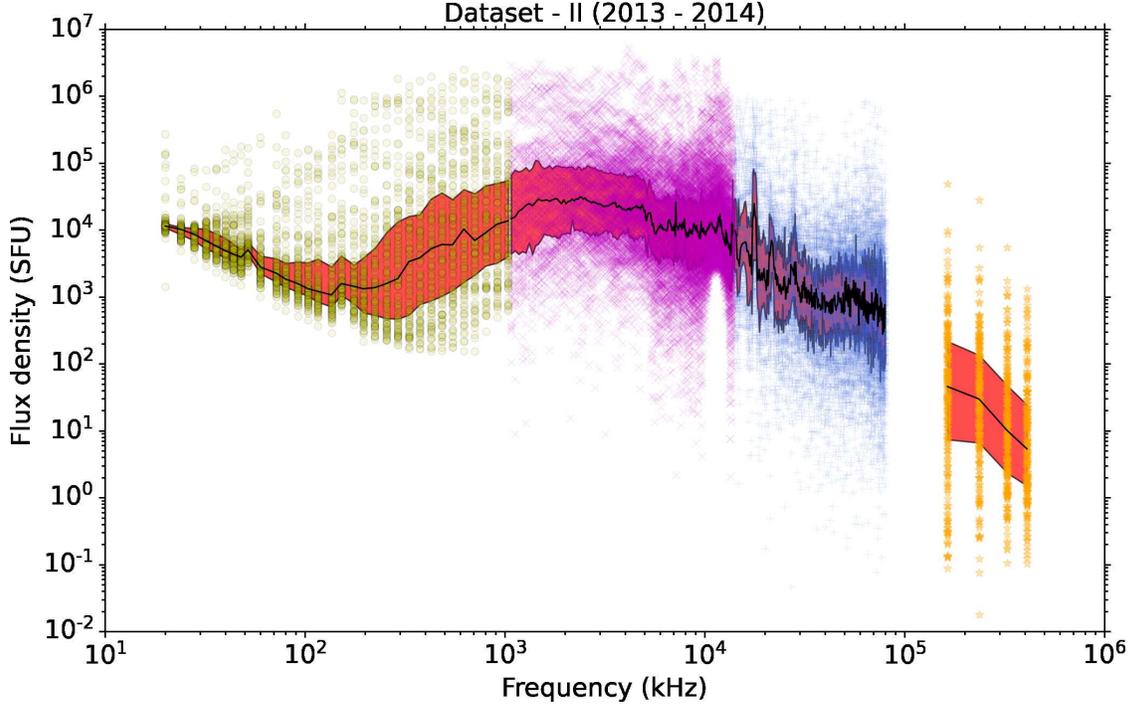}}
\caption{Spectral response of type III bursts observed from 20 kHz to 410 MHz is shown. The yellow `$\circ$', magenta `$\times$', and blue `$+$' markers indicate the maximum flux density (in SFU) of the 115 type III bursts observed using RAD1 and RAD2 receivers of Wind/Waves instrument and Nan\c{c}ay Decameter Array respectively. The orange `*' markers indicate the observations of Nan\c{c}ay Radioheliograph at selected frequencies (i.e., 164, 236, 327, 410 MHz). The black solid curve represents the median (\emph{i.e.}, 50th percentile) of the flux densities of the data shown in every frequency channel. Red colored filled region shows the first and third quartiles of the data shown. Note that the galactic contribution is subtracted as described in Section \ref{sec:observations}.}
\label{fig:nda_wind115}
\end{figure}

Further, we have studied the peak flux densities of the 115 type III bursts of dataset-II. The outcome is displayed in Figure~\ref{fig:nda_wind115}. The data points in yellow `$\circ$', magenta `$\times$' and blue `$+$' correspond the Wind/WAVES/RAD1, Wind/WAVES/RAD2, and the NDA, respectively. Again, the black solid curve represents the medians of the data points for all the events as a function of the frequency, and the filled red color region indicates the first and third quartiles. Two remarks can be made on this Figure. Firstly, the spectral profile of the median Type III flux displayed in Figure~\ref{fig:nda_wind115} is globally continuous between Wind/WAVES and the NDA, suggesting that our calibration techniques are reliable. 
Secondly, as for dataset-I, and despite fewer events, the median spectral profile of dataset-II also exhibits a maximum at about 1.5 MHz with a maximum flux of about $3 \times 10^4$ SFU. Even though it presents a more substantial variability as a function of frequency, the median spectral profile globally decreases from $3 \times 10^4$ SFU at 1.5 MHz to about $5\times10^2$ SFU at 80 MHz. 

The more significant fluctuations in the 5 to 80 MHz frequency range can be attributed to (i) the resonance frequency of the half-wave dipole Wind/Waves antenna, (ii) RFI from the spacecraft, (iii) the strong terrestrial radio frequency interference (RFI), and (iv) the ionospheric absorption. Indeed a relative decrease in the flux density is seen from 10 to 40 MHz in the Figure~\ref{fig:nda_wind115}. 

In addition to the Wind/Waves and the NDA data, we have over-plotted in Figure~\ref{fig:nda_wind115} the peak flux densities of type III bursts measured using imaging observations of the NRH, i.e., dataset-III. These data are displayed by orange `$*$' markers. As before, the black solid curve in the 150 - 450 MHz range represents the medians of the NRH data points for all the events, and the filled reddish color region indicates the first and third quartiles. As for the NRH frequency range, the median type III radio flux decreases as a frequency function in the NRH range. As datasets II and III do not correspond to the same statistical samples of solar type III radio bursts, we can consider that these two samples are representative of the mean evolution of the radio power of types III bursts with the frequency. Therefore we show, for the first time, that the typical type III radio flux decreases continuously from $3\times10^4$ SFU at 1.5 MHz to about 5 SFU at 400 MHz.

\subsection{Summary of our statistical analysis}

Before presenting the summary of our analysis and displaying the profiles of the medians of the peak fluxes, we have to remove the Quasi-Thermal Noise's (QTN) contribution at low frequencies and discuss the uncertainties of our measurements. Above the plasma frequency and for an antenna of length much longer than the local Debye length, which is the case for the WIND spacecraft, the QTN varies as $f^{-3}$ for $f \gg f_p$ \citep{MV2000,Nicole1989}. For frequencies $f \gtrsim f_p$ the variation is less steep and is roughly $f^{-0.5}$ in the case of WIND. At one AU, the plasma frequency typically varies between 10 and 40 kHz. Therefore, it is reasonable to assume that for the 20 to 150 kHz range (where we see the QTN in our data; see Figure \ref{fig:wind2000} and \ref{fig:nda_wind115}), QTN varies as a power law with a spectral index between the two cases mentioned above. Therefore instead of removing the QTN from each single spectrum, we have subtracted by fitting the median data curves below 150 kHz with the following empirical power-law model,

\begin{equation}
S_\text{SFU} = S_0 \times (f/f_0)^{\alpha}\,
\end{equation}

where $S_\text{SFU}$ is the flux density in SFU, $S_0$ is the flux density corresponding to the starting frequency of observations (i.e., $f_0 = 20$ kHz), f is the frequency of observations below 150 kHz, and $\alpha$ is the power-law index.  The power-law index we obtain for both datasets is $\alpha \approx -1.3$, which is well in the expected range [-3,-0.5]. Figure \ref{fig:comparison} shows the spectral response of the type III bursts after QTN is removed.

\begin{figure}[!ht]
   \centerline{\hspace*{0.015\textwidth}
               \includegraphics[width=0.49\textwidth,clip=]{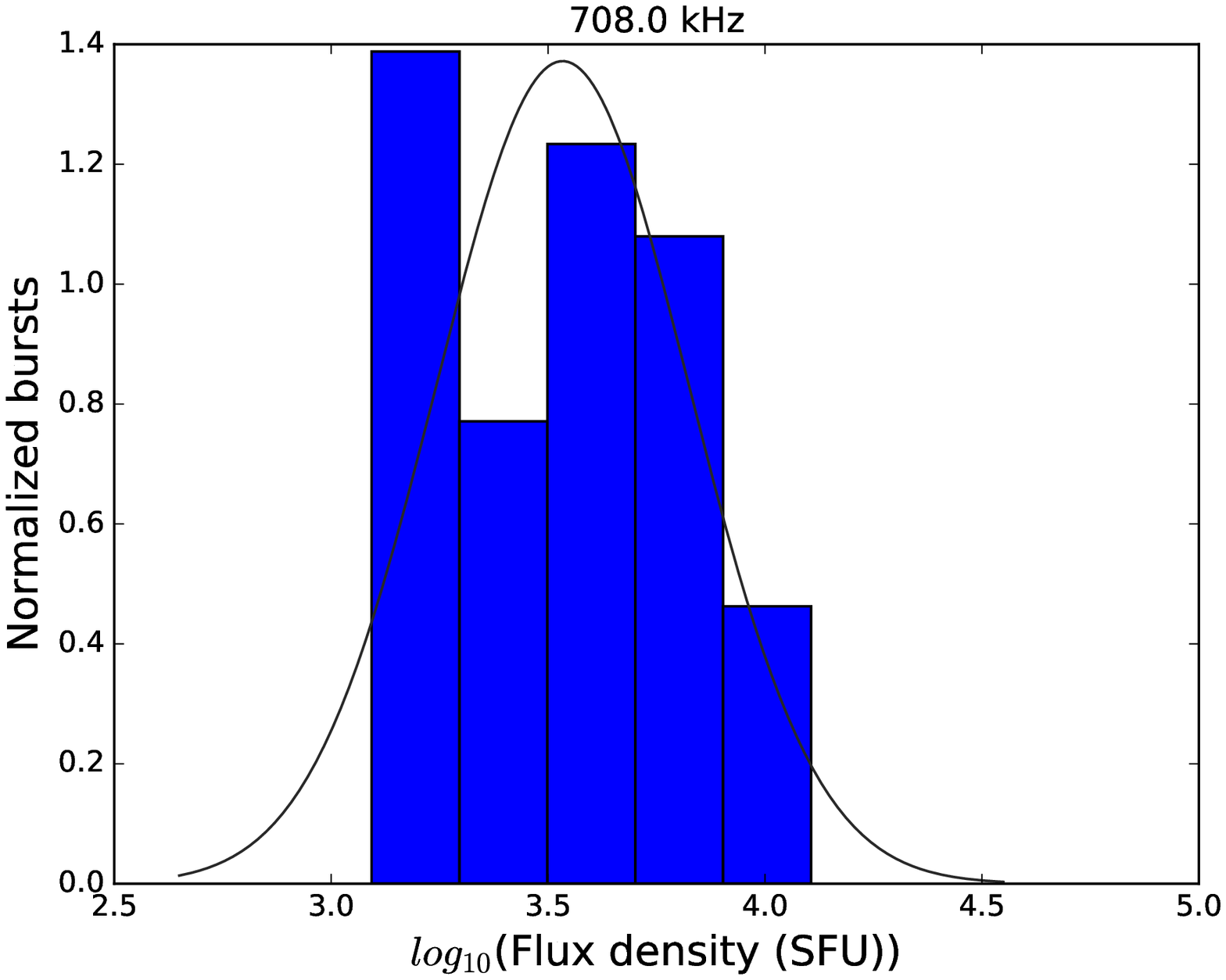}
               \includegraphics[width=0.49\textwidth,clip=]{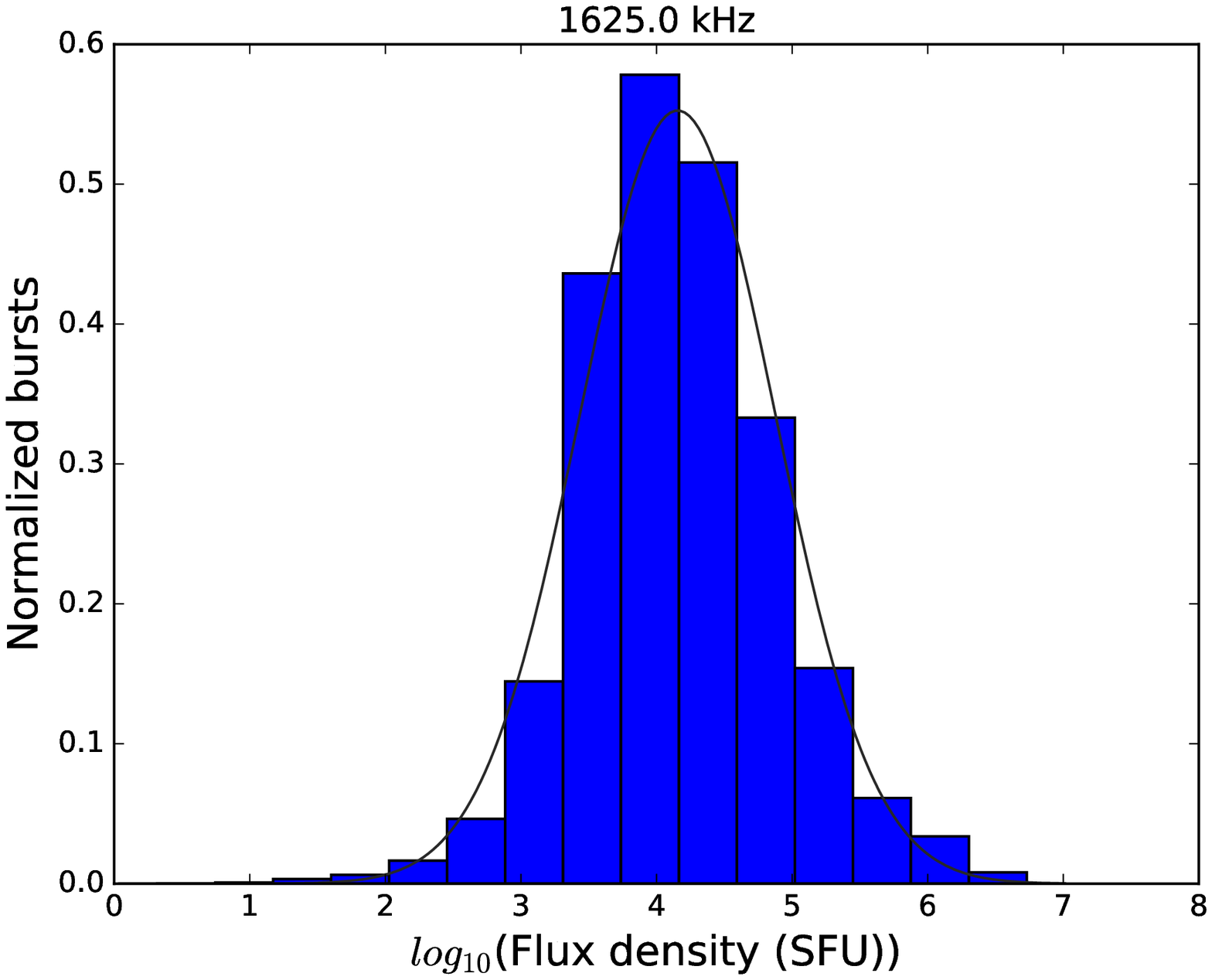}
              }
     \vspace{-0.375\textwidth}   
     \centerline{\Large \bf     
      \hspace{0.075 \textwidth}  \color{black}{(a)}
      \hspace{0.43\textwidth}  \color{black}{(b)}
         \hfill}
     \vspace{0.33\textwidth}    
   \centerline{\hspace*{0.015\textwidth}
               \includegraphics[width=0.49\textwidth,clip=]{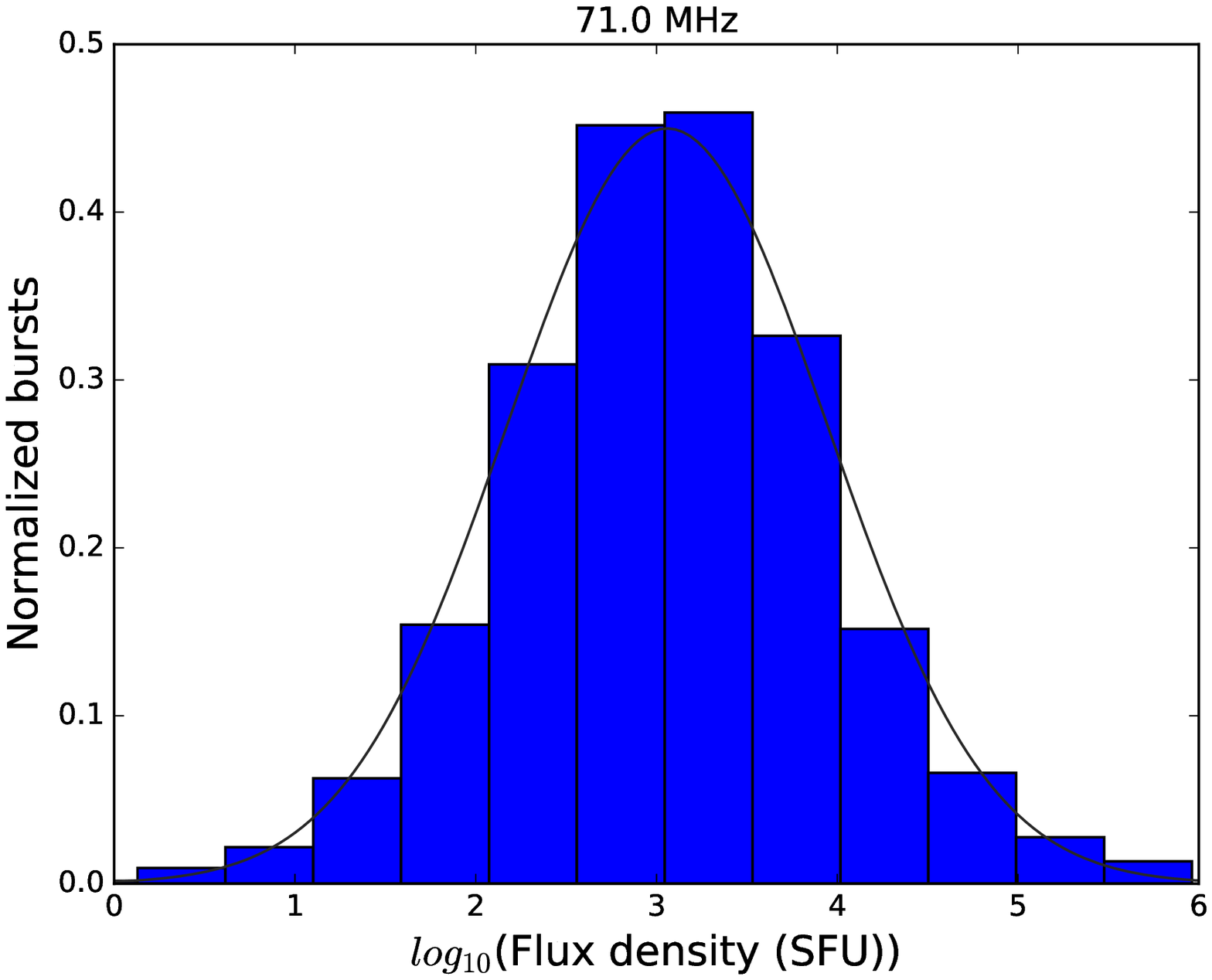}
               \includegraphics[width=0.49\textwidth,clip=]{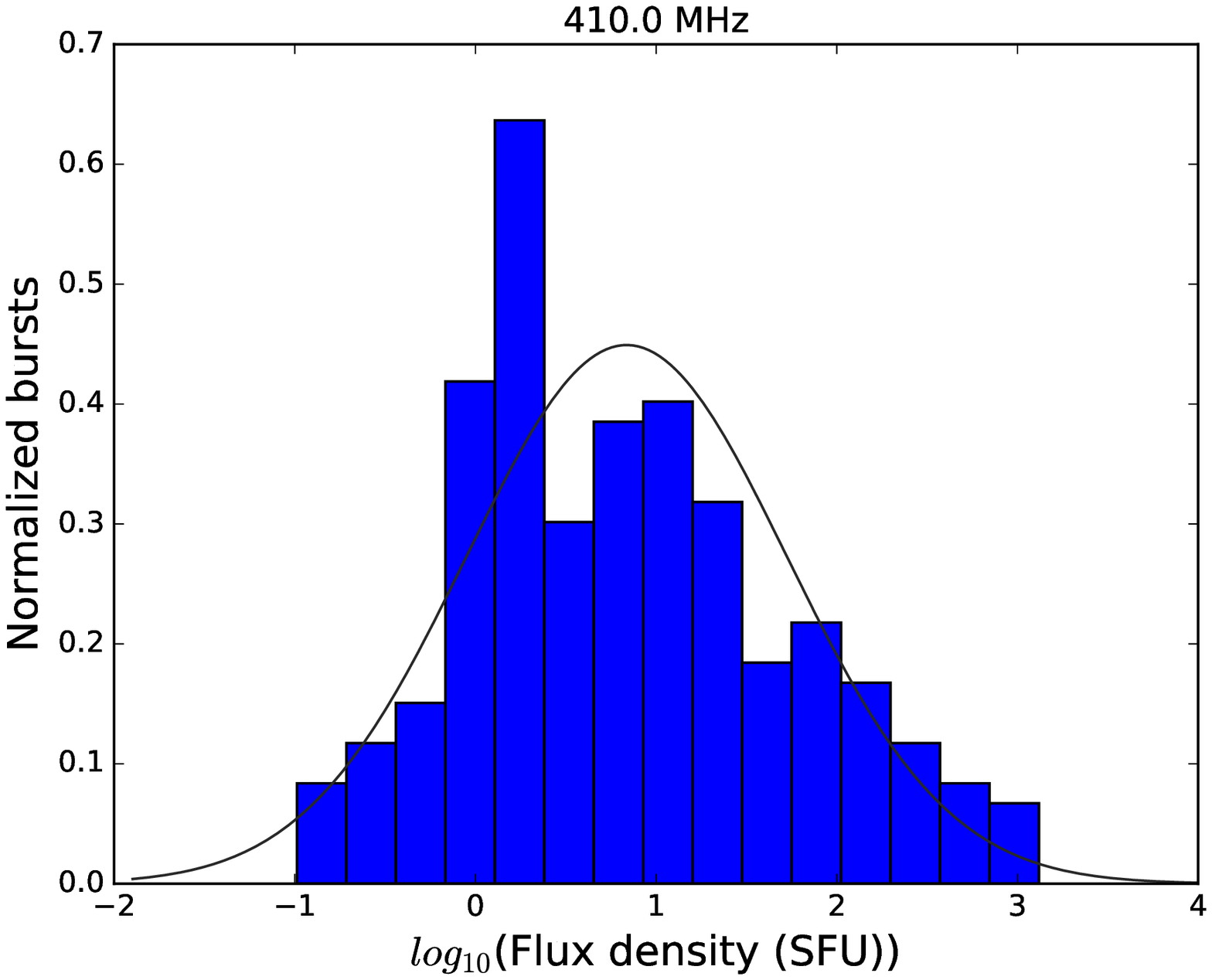}
              }
     \vspace{-0.375\textwidth}   
     \centerline{\Large \bf     
      \hspace{0.075\textwidth} \color{black}{(c)}
      \hspace{0.43\textwidth}  \color{black}{(d)}
         \hfill}
     \vspace{0.33\textwidth}    
   
\caption{Sub-panels (a), (b), (c), and (d) show the log-normal distribution of flux density of Type III bursts observed by RAD1, RAD2, NDA, and NRH, respectively. A black curve indicates the normal distribution of the logarithm of flux densities.}
\label{fig:log_normal}
\end{figure}

\begin{figure}[!ht]
\centerline{\includegraphics[width=20cm]{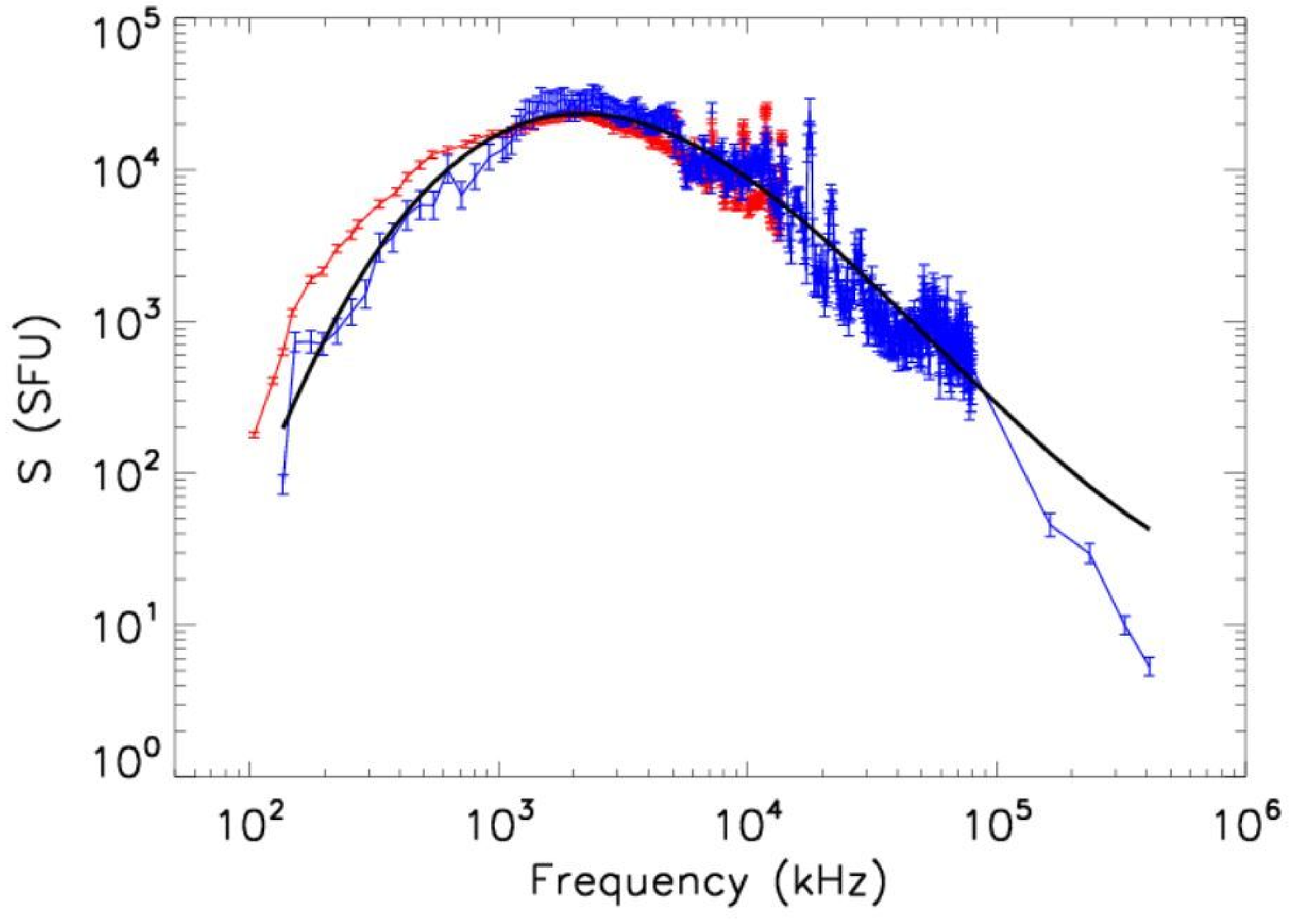}}
\caption{Median curves from Figures \ref{fig:wind2000} (derived using dataset-I in red) and \ref{fig:nda_wind115} (derived using dataset-II and III in blue) after removing the quasi-thermal noise detected by the antennas below 150 kHz and computing the error bars. The black solid line represents the empirical polynomial model of $Y=\ln{(S_\text{SFU})}$ as a function of $X=\ln{(F_\text{MHz})}$ (see text for more details).}
\label{fig:comparison}
\end{figure}

Concerning the uncertainties of our measurements, we have made the following assumption. For an observable $x$ which is normally distributed, the uncertainty on the mean value $\langle x\rangle$ of a sample of N measurements of $x_i$ is $\sigma_x/\sqrt{N}$, where $\sigma_x$ is the standard deviation of all the $x_i$. However, our radio fluxes $S_i$ are rather represented by log-normal distributions and not by normal ones. The distributions of log~$(S_i)$ are close to normal as shown in sub-panels of Figure~\ref{fig:log_normal}. The sub-panels show the RAD1 (708.0 kHz), RAD2 (1625 kHz), NDA (71 MHz), and NRH (410 MHz) observations. Also, this can be seen in Figures \ref{fig:wind2000} and \ref{fig:nda_wind115} where the filled red color regions, indicating the first and third quartiles are symmetric with respect to the median values. Furthermore, the medians and means of the distribution of log~$(S_i)$ are similar, which is not the case of the medians and means of $S_i$. Thus the error bars must be computed over $log(S _i)$ distributions.

Figure~\ref{fig:comparison} presents, therefore, the summary of our statistical analysis. It displays the median radio fluxes profiles of both dataset-I (Figure~\ref{fig:wind2000}) in red and dataset-II and III (Figure~\ref{fig:nda_wind115}) in blue. For these profiles, we have removed the QTN at low frequencies and added the error bars in log scale, defined as $\sigma_{log(S)}/\sqrt{N}$, where $\sigma_{log(S)}$ are the standard deviations of log~$(S_i)$ for all frequencies. For the Wind/Waves frequency range, as the standard deviations are similar for dataset-I and II, the error bars are smaller for dataset-I because of the larger number of samples (1434 instead of 115).

As indicated in Table 1, the three datasets do not correspond to the same period of the solar cycle. Dataset-I covers observations from 1995 to 2009, which corresponds to the full solar cycle 23, while dataset-II covers a period of solar cycle 24 restricted to only the maximum of this cycle, which occurred in mid-2014. As for dataset-III is concerned, it covers a period spanning from the declining phase of solar cycle 23 up to the maximum of solar cycle 24. These different solar cycle coverages suggest that, whatever the observation period, as long as the number of Type III considered is large enough, we observe the same behavior of the maximum of the radio flux as a function of the frequency. Namely, the maximum of radio emission at about 2 MHz is an intrinsic property of the Type III generation at decametric to kilometric wavelengths. This property seems to be confirmed even in the metric range since the radio fluxes of dataset-III are also decreasing as a function of the frequency and are well below the fluxes observed by the NDA in the decametric range. We have therefore confirmed and precised the initial observations by  \citet{Web1978}, \citet{Bon2008b} and \citet{Krupar2014}, by extending them for the first time up to the metric frequency range. In order to quantify this main result and make it usable for future theoretical studies, we have fitted the data from the dataset II and III with a three-order polynomial model and obtained that the empirical model which best represents the observations is 
\begin{equation}\label{eq:pol}
Y = 0.04 X^3 - 1.63 X^2 + 16.30 X -41.24
\end{equation}

\noindent with $X=\ln{(F_\text{MHz})}$ and  $Y=\ln{(S_\text{SFU})}$. T
his model is represented in Figure~\ref{fig:comparison} by the black solid line.

A final remark concerning Figure~\ref{fig:comparison} is that the blue and red curves overlap, within the error bars, at all frequencies above 1 MHz. Below 1 MHz, the radio fluxes are slightly lower for dataset-II than for dataset-I, even though the global evolution is the same, namely a decrease of the flux as a function of the decreasing frequency in this range. This behavior is a bit puzzling and is under investigation.

\subsection{Possible explanations for the maximum of the radio flux density of type III bursts}

\begin{figure}[!ht]
\centerline{\includegraphics[width=15cm]{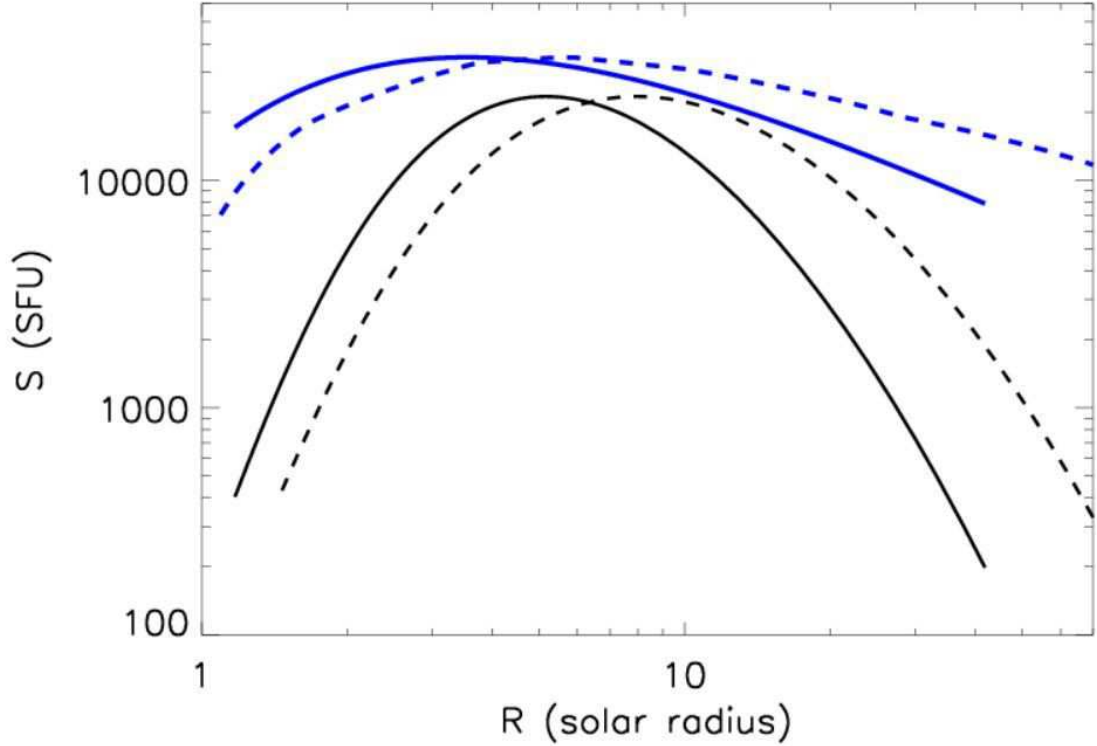}}
\caption{The same median fluxes as those shown in Figure~\ref{fig:comparison} are displayed as a function of the distance from the Sun. Here we use the Sittler and Guhathakurta model for coronal streamers \citep{Sit1999} and assume a fundamental emission for type III bursts (black solid lien) or the harmonic (black dashed line). The solid and dashed lines represent tentative to explain the median radio flux (see text for more details).}
\label{fig:discussion}
\end{figure}

Assuming a density model for the corona and solar wind, it is possible to locate the maximum of the radio flux density of type III bursts. Using the Sittler and Guhathakurta model for coronal streamers \citep{Sit1999} and assuming a fundamental emission for type III bursts, the black solid line in Figure~\ref{fig:discussion} displays the previously found empirical model of the radio radio flux as function of the distance from the Sun and given by equation \ref{eq:pol}.

Note, however, that there is a possibility that the maximum of Type III emissions we observe is a combination of both fundamental and harmonic emissions \citet{Krupar2014}. The black dashed line in Figure~\ref{fig:discussion} represents the same empirical model for the radio flux, assuming an emission at the harmonic of the plasma frequency for the Type IIIs. The maximum of radio power falls therefore in the  range 4 to 10 $R_{\odot}$ (from center of the Sun) depending whether the emission is assumed to be at the fundamental or the harmonic. This radial range is crucial for the solar wind since this is where it becomes both supersonic, after the Parker sonic point, and super-Alfv\'{e}nic. 

\citet{Krupar2014} have proposed a simple model in which the radio emission is at harmonic and saturated with a radio flux density proportional to the energy of Langmuir waves \citep{Mel1980}. Assuming a fast relaxation of the energetic electron beam as in \citep{Kont2001}, \citet{Krupar2014} obtain that the radio flux is simply proportional to $1/(r^{2}f_p)$ (see their equation 12). Figure~\ref{fig:discussion} shows the \citet{Krupar2014} model by a blue solid line. As can be seen, the maximum of the radio flux for this model is somewhat below the maximum of the observed radio fluxes.

Given the density fluctuations could affect the growth of Langmuir waves, the Type III radio generation will be influenced by the density fluctuations in the corona and solar wind. This in turn, could affect the location of the maximum of the observed radio fluxes. 
Specifically, the generation of Langmuir waves could be suppressed by higher level of density fluctuations, so the maximum is shifted 
with respect to the simplified model by \citet{Krupar2014}.
There are no in situ observations of the density fluctuations in the outer corona. 
The only current possibility is to model and/or simulate the evolution of the plasma turbulence in this region and to assume that the density fluctuations have a suitable trend. An additional complication connected with density turbulence is the scattering of radio waves between the source and observer. The scattering broadens the burst time profile and hence can reduce the peak flux detected at 1 AU.

Nevertheless a full comprehensive model which could explain in details why there is a maximum of the Type III radio power around 5-6 $R_{\odot}$ is still to be found.

\section{Summary and Conclusions}\label{sec:summary}

In this article, we have performed the first ever statistical analysis of the spectral response of solar type III bursts over the wide frequency range between 20 kHz and 410 MHz. 
For this purpose, we have used the observations that were carried out using both spaced-based (Wind/Waves) and ground-based (Nan\c{c}ay Decameter Array and Nan\c{c}ay Radioheliograph) facilities. In order to compare the flux densities observed by the different instruments, we have carefully calibrated the data and displayed them in Solar Flux Units. We have studied 1434 type III bursts that were observed by  Wind/Waves solely (see ~\ref{fig:wind2000}) and 115 type III bursts that were observed simultaneously by both Wind/Waves and NDA (see Figure~\ref{fig:nda_wind115}). Note that in the latter case, both instruments observed type III bursts which were roughly in the same line-of-the-sight. In addition, we have included the data derived using calibrated observations of the NRH at some particular frequencies. 

The main results of our study is that type III bursts in the metric to hectometric wavelengths range statistically exhibit a maximum of their radio power at around 1 to 2 MHz. Using the Sittler and Guhathakurtha model for coronal streamers \citep{Sit1999}, we have found that this frequency range corresponds to the heliocentric distance range $\approx 3-8$ $R_{\odot}$ if one assume a radio emission at the fundamental plasma frequency. On the other hand if the type III bursts are a combination of both fundamental and harmonic emissions, the maximum of the radio flux would originate in the heliocentric distance
range of $\approx$ 1 to 20 $R_{\odot}$. This radial range is crucial for the solar wind since this is where it becomes both supersonic, after the Parker sonic point, and super-Alf\'{v}enic. 

Finally there is no doubt that our findings will soon be compared to in-situ observations. Indeed the Parker Solar Probe \citep{Fox2016} is expected to reach 10 $R_{\odot}$ in 2024. The in-situ measurements of the various plasma parameters including density fluctuations, Langmuir waves and energetic electrons should provide crucial information for fully explaining the maximum radio flux of Solar type III bursts.


\acknowledgments
KSR acknowledges the financial support from the Centre National d'\'{e}tudes Spatiales (CNES), France. V. Krupar acknowledges the support by NASA under grants 18-2HSWO218\_2-0010 and 19-HSR-19Z\_2-0143. The authors acknowledge the Nan\c{c}ay Radio Observatory / Unit\'e Scientifique de Nan\c{c}ay of the Observatoire de Paris (USR 704-CNRS, supported by Universit\'e d'Orl\'eans, OSUC, and R\'egion Centre in France) for providing access to NDA observations accessible online at \url{http://www.obs-nancay.fr}.
French coauthors were supported by the CNRS/INSU PNST program. We thank the referee for providing useful suggestions that helped in improving the manuscript.




\bibliographystyle{aasjournal}
\bibliography{ms}



\end{document}